\documentclass[draft]{agujournal2019}

\draftfalse
\journalname{Radio Science}

\begin{document}
    \title{The Development of Spatial Attention U-Net for The Recovery of Ionospheric Measurements and The Extraction of Ionospheric Parameters}
    \authors{Guan-Han~Huang\affil{1}, Alexei~V.~Dmitriev\affil{1,2} and Chia-Hsien~Lin\affil{1}, Yu-Chi~Chang\affil{1}, Mon-Chai~Hsieh\affil{1}, Enkhtuya Tsogtbaatar\affil{3}, Merlin~M.~Mendoza\affil{1}, Hao-Wei~Hsu\affil{1}, Yu-Chiang~Lin\affil{1}, Lung-Chih~Tsai\affil{4}, Yung-Hui~Li\affil{5}}
    \affiliation{1}{Department of Space Science and Engineering, National Central University, Taoyuan City 320317, Taiwan}
    \affiliation{2}{Skobeltsyn Institute of Nuclear Physics, Lomonosov Moscow State University, 119899 Moscow, Russia}
    \affiliation{3}{Department of Computer Science and Information Engineering, National Central University, Taoyuan City 320317, Taiwan}
    \affiliation{4}{Center for Space and Remote Sensing Research, National Central University, Taoyuan City 320317, Taiwan}
    \affiliation{5}{AI Research Center, Hon Hai Research Institute, Taipei 114699, Taiwan}

    \correspondingauthor{Guan-Han~Huang}{enter468@g.ncu.edu.tw}

    \begin{keypoints}
    \item A deep learning model is applied to the ionogram recovery.
    \item The model can well identify the combined signals of the sporadic E
    layer and the ordinary and extraordinary modes of the F2 layer signal.
    \item Critical frequencies of the modes, and the intersection frequency
    between them are derived.
    \end{keypoints}

    \begin{abstract}
    We train a deep learning artificial neural network model,
    Spatial Attention U-Net to recover useful ionospheric signals from noisy
    ionogram data measured by Hualien's Vertical Incidence Pulsed Ionospheric
    Radar.
    Our results show that the model can well identify F2 layer ordinary and 
    extraordinary modes (F2o, F2x) and the combined signals of the E layer
    (ordinary and extraordinary modes and sporadic Es).
    The model is also capable of identifying some signals that
    were not labeled.
    The performance of the model can be significantly degraded by
    insufficient number of samples in the data set.
    From the recovered signals, we determine the critical frequencies of
    F2o and F2x and the intersection frequency between the two signals.
    The difference between the two critical frequencies is peaking at 0.63 MHz, 
    with the uncertainty being 0.18 MHz.
    \end{abstract}

    \section*{Plain Language Summary}
    A large amount of images are retrieved by a specialized instrument designed
    to make observations in the ionosphere.
    These images are contaminated by instrumental noises.
    In order to recover useful signals from these noises, we train a deep
    learning model.
    A dataset containing the labeled signals are used for both training and 
    validating the model performance.
    The desired signals are manually labeled using a labelling software.
    By comparing the model predictions with the labels, the
    results show that the model can well-identify the elongated, overlapping, 
    or compact signals.
    The model is also capable of correcting some missing and incorrect labels.
    The performance of the model is sensitive to the data number of 
    the corresponding labels fed to the model during training.
    The recovered useful signals are then used to estimate physical quantities
    which are important for the study of ionospheric physics.

    \section{Introduction}
    The ionosphere is a region of ionized gases, plasmas,
    populating the upper atmosphere and thermosphere \cite{intro}.
    The ionosphere consists of layers concentrated at specific heights.
    Radio waves propagate through the ionospheric layers at different group 
    velocities and, hence, split into different wave modes
    according to the electron density, the magnetic field, etc.
    An experimental ground-based technique of ionosondes 
    has been used for a long time to investigate the vertical profile of the 
    ionospheric ionization represented by the density of free electrons, 
    so-called electron content (EC).

    The data product of ionosonde measurements are ionograms,
    which exhibit signals deflected by the ionosphere at various
    virtual heights as a function of the sounding frequency.
    The virtual height of the deflection is obtained by assuming that the wave 
    beams are propagating at the speed of light.
    The sounding frequency at which the virtual height rapidly increases is
    called the critical frequency, which also corresponds to the local
    maximum of the EC.
    In addition, the splitting in the sounding frequency between the signals
    of different wave modes is related to the local magnetic field.
    These ionogram parameters can be used for
    the true height analysis \cite{true_height_1,true_height_2},
    estimating the magnetic field strength \cite{handbook}, and modelling
    the electron density higher than the deflection height by the Chapman
    function \cite{topside}.
    Furthermore, the stability of some ionogram interpretation
    algorithms \cite{oxpoint_a,oxpoint_b} rely on the intersection point of
    the ordinary mode and the extraordinary mode signals.

    The ionograms from Hualien's
    Vertical Incidence Pulsed Ionospheric Radar (VIPIR) are
    featured by small and compact signals or thin and elongated signals.
    These ionograms are contaminated by stripe noises
    appearing in many frequency bands.
    \citeA{snr} showed that the Hualien dataset has stronger
    noise signals compared with the Jicarmaca dataset \cite{aeperu}.

    Thousands of measurements are produced by VIPIR per day.
    It is hard work and time-consuming for skilled researchers
    to recover ionospheric signals from such immense dataset.
    Therefore an automated method based on fuzzy logic \cite{fuzzy}
    has been applied.
    In recent years, there are also deep learning techniques
    applied to the ionogram recovery \cite{unetseg,dias,aeperu}.

    In this research, we implement a deep learning model to the
    Hualien VIPIR ionograms, and recover different ionogram signals.
    The data and the preprocessing of the dataset are presented in
    Section~\ref{sec:data}.
    The deep learning model and the validation of the model are
    described in Section~\ref{sec:model}.
    In Section~\ref{sec:result}, 
    we evaluate the performance of signal recovery (in \ref{sec:recover}),
    and derive the ionogram parameters from the
    recovered ionograms (in \ref{sec:param}).
    The results are discussed in Section~\ref{sec:discuss}
    and summarized in Section~\ref{sec:summary}.

    \section{Data}
    \label{sec:data}
    We use the ionograms acquired from the Hualien VIPIR digisonde
    operated at Hualien, Taiwan ($23.8973^\circ$N, $121.5503^\circ$E).
    The dataset contains $6131$ ionograms spanning from
    2013/11/08 to 2014/06/29.
    Each ionogram covers a virtual height range up to 800km,
    and sounding frequency range from 1MHz to 22MHz,
    and the signal amplitude range up to 100 decibels (dB).

    To reduce the data size and remove the calibration signals,
    we reduce the size of ionograms to virtual height range from 66 to 600km,
    and sounding frequency from 1.58 to 20.25MHz.
    Different useful signals in each ionogram are manually identified and
    labeled into polygons using \texttt{labelme} \cite{labelme}.
    The polygons are rasterized into a binary array of dimension 800x1600x11,
    which correspond, respectively, to frequency, height and the label.
    The eleven labels of useful signals are defined as follows
    (see Figure~\ref{fig:labeling}):

    \begin{figure}
        \centering
        \includegraphics[width=.8\textwidth,clip,trim=0 0 0 0]{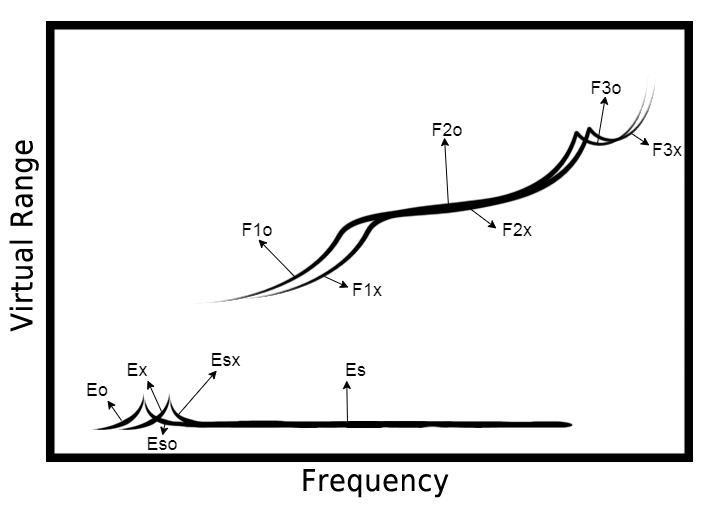}
        \caption{A schematic diagram of ionospheric signals. Different signals are indicated by arrows with annotations (see the text for details).}
        \label{fig:labeling}
    \end{figure}

\begin{figure}
        \includegraphics[width=\textwidth,clip,trim=0 2cm 0 2cm]{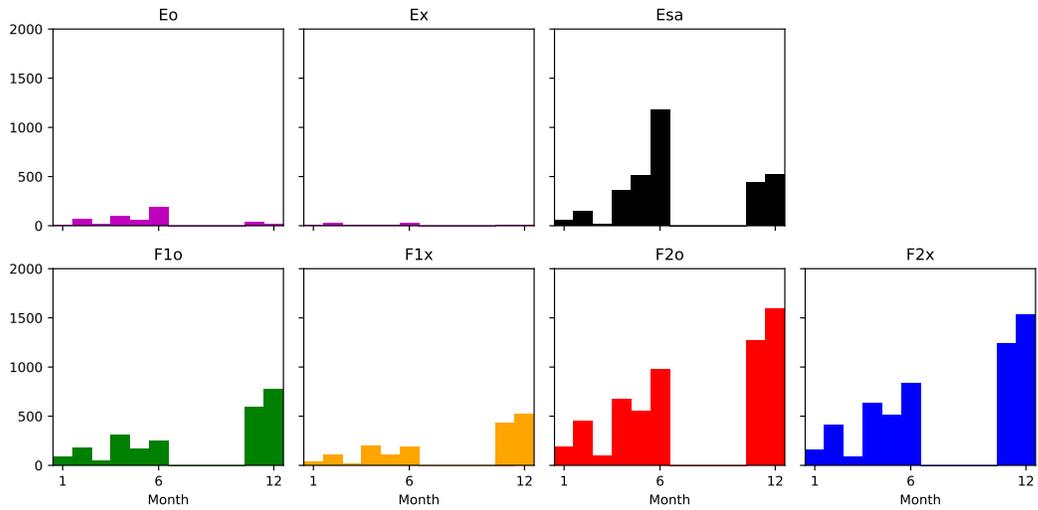}
        \caption{The panels in (a) and in (b) are the local time distributions 
        and seasonal distributions, respectively, of the 7 signal labels, as
        indicated above the corresponding panels. All panels follow the scale 
        of the left most y-axis.}
        \label{fig:lt_mm_7lab}
    \end{figure}

    \begin{enumerate}
        \item Eo: Ordinary signal of the E-layer, with the virtual height increasing as the sounding frequency increased.
        \item Ex: Extra-ordinary signal of the E-layer, with the virtual height increasing as the sounding frequency increased.
        \item Eso: Ordinary signal of the sporadic E-layer, with the virtual height decreasing as the sounding frequency increased.
        \item Esx: Extra-ordinary signal of the sporadic E-layer, with the virtual height decreasing as the sounding frequency increased.
        \item Es: Signal of the sporadic E-layer, with the virtual height constant as the sounding frequency increased.
        \item F1o: Ordinary signal of the F1-layer, with the virtual height increasing as the sounding frequency increased.
        \item F1x: Extra-ordinary signal of the F1-layer, with the virtual height increasing as the sounding frequency increased.
        \item F2o: Ordinary signal of the F2-layer.
        \item F2x: Extra-ordinary signal of the F2-layer.
        \item F3o: Ordinary signal of the F3-layer, with the virtual height increasing as the sounding frequency increased.
        \item F3x: Extra-ordinary signal of the F3-layer, with the virtual height increasing as the sounding frequency increased.
    \end{enumerate}
    The $11$ signals shown in Figure~\ref{fig:labeling} usually do not appear
    simultaneously in all ionograms, 
    and some signals can be too faint to be labeled.
    The percentages of different labeled signals in our data set are
    7.89\% for Eo, 0.99\% for Ex,
    39.81\% for Es, 13.34\% for Eso, 10.59\% for Esx,
    39.19\% for F1o, 26.06\% for F1x,
    94.65\% for F2o, 88.90\% for F2x,
    0.13\% for F3o and 0.16\% for F3x.
    Eso, Esx, F3o, and F3x have very poor statistics.
    Therefore, we discarded F3o and F3x to increase the statistics of E layer,
    and combine Eso, Esx, and Es labels into the Esa label.
    As a result, we reduce $11$ labels into $7$ labels.
    The panels in Figure~\ref{fig:lt_mm_7lab}a show the local time 
    distribution of the occurrence of the $7$ labels in the bulge of the 
    equatorial ionoization anomaly at Taiwan.
    During the studied time in Taiwan, Esa, F2o, and F2x labels have the 
    highest occurrence rate, and Eo and Ex the lowest.
    Sporadic Esa and F2 (F2o and F2x) layers occur at all dayside
    local time hours.
    F1 layer (F1o and F1x) does not occur in the evening.
    The E layer (Eo and Ex) occurs mainly in the morning.
    The panels in Figure~\ref{fig:lt_mm_7lab}b show the seasonal distribution 
    of the occurrence of the layers.
    F2o and F2x as well as F1o and F1x appear in all seasons and have similar 
    distributions.
    The sporadic layer Esa has the highest occurrence rate during
    the summer \cite{e_summer}.
    Our focus is to recover the Esa, F2o, and F2x labels since the radio wave 
    propagation in the Taiwan region is mostly affected by these layers,
    due to their high occurrence rate.

    Finally, the dataset was split into ratios of 64\%, 16\% and 20\%,
    respectively, for the training set, the validation set and the test set,
    resulting in 3925, 981, 1226 ionograms in each set.
    The coverage ratios of the seven labels in each set are
    shown in Table~\ref{tab:datacov}.

    \begin{table}
        \centering
        \begin{tabular}{*{8}{c}}
            \hline
             & Eo & Ex & Esa & F1o & F1x & F2o & F2x \\
            \hline
            Train (\%) & 8.13 & 0.97 & 52.60 & 39.22 & 26.45 & 94.37 & 88.10 \\
            Validation (\%) & 8.36 & 1.33 & 53.31 & 38.12 & 22.53 & 95.82 & 87.77 \\
            Test (\%) & 6.77 & 0.82 & 52.04 & 39.97 & 27.65 & 94.62 & 88.34 \\
            \hline
        \end{tabular}
        \caption{Coverage ratio of each signal label in percentage for 
        training, validation and testing set.}
        \label{tab:datacov}
    \end{table}

    \section{Methodology}
    \label{sec:model}
    \subsection{Deep Learning Model}
    The model employed for this study is Spatial-Attention U-Net (SA-UNet),
    developed by \citeA{SDUNet,SAUNet}.
    The SA-UNet is featured by a U-Net \cite{unet} architecture with a spatial
    attention module at the bottle-neck of the model structure.
    U-Net has been successful in classifying an image into different labels.
    With modifications to the U-Net, SA-UNet has been shown that it is capable 
    of identifying vessels from the eyeball images \cite{SAUNet}.
    The model implementation is also available on
    Github (\url{https://github.com/clguo/SA-UNet}).
    Since vessels and ionogram traces are both tiny or elongated features,
    we consider that SA-UNet is suitable for the ionogram recovery.

\begin{figure}
        \centering
        \includegraphics[width=\textwidth,clip,trim=0 10cm 0 0]{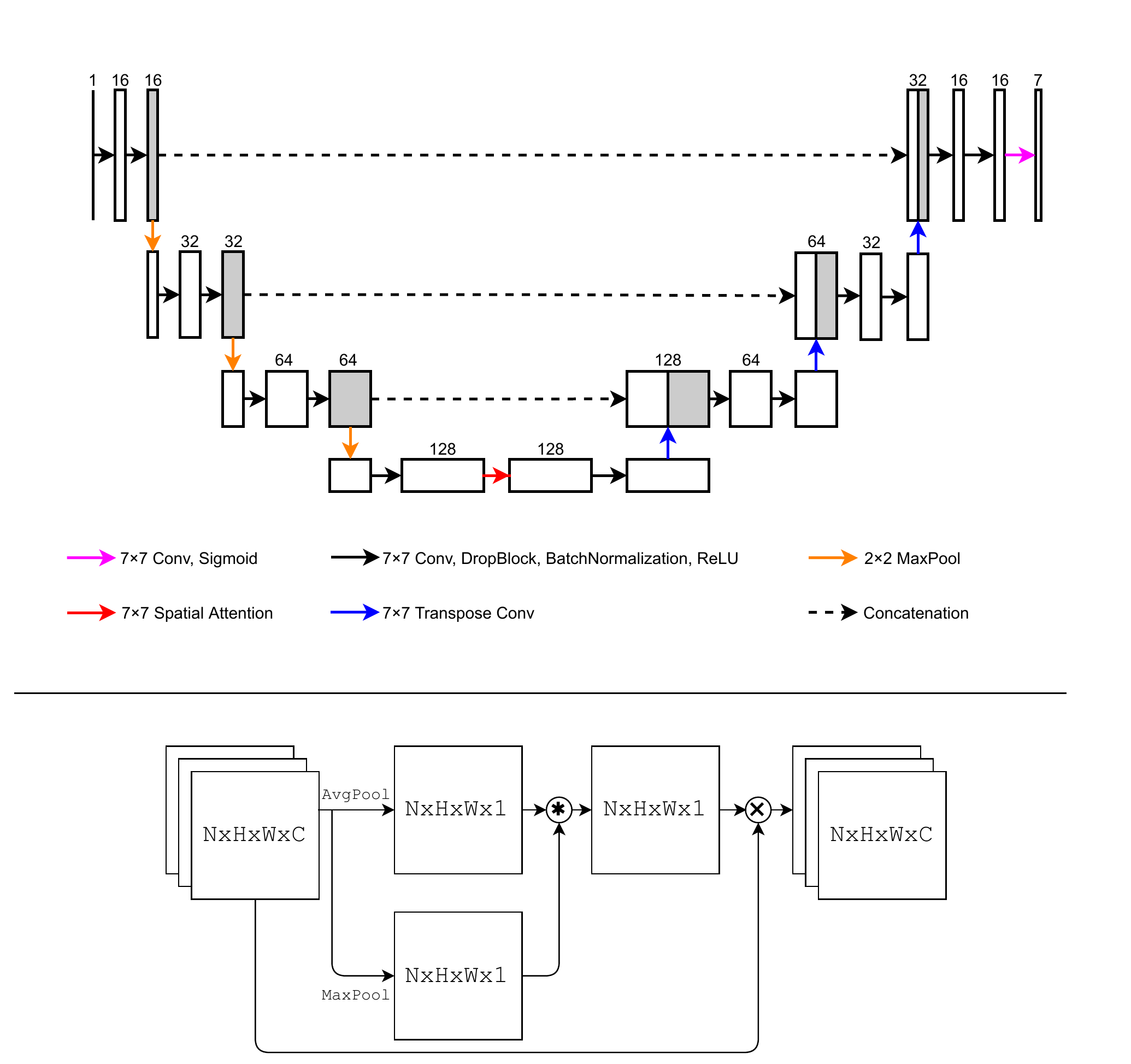}
        \caption{The upper panel shows the architecture of the SA-UNet.
        The number on top of each rectangle (layer) indicates the 
        dimensionality along the feature space of the input or the output.
        The lower panel shows the structure of the Spatial Attention module,
        which is the red arrow in the upper panel.
        \texttt{N}, \texttt{H}, \texttt{W}, \texttt{C} denote the 
        dimensionality along the sample, height, width, and feature space, 
        respectively.
        The asterisk and cross signs represent the convolution and the
        product of array elements, respectively.}
        \label{fig:sa_unet}
    \end{figure}

    The architecture of the SA-UNet is shown in Figure~\ref{fig:sa_unet}.
    When an ionogram is sent through the model,
    the convolution block extracts the features,
    and the pooling layer reduce the image size.
    The DropBlock \cite{DropBlock1} randomly drops the image pixels to
    virtually increase the sample size, which can potentially reduce the
    overfitting.
    The spatial attention module rescales the features,
    so that the model can put more emphasis on important features.
    The features extracted are assembled and localized in the
    transpose convolution blocks.
    The skip-connections return the image size back to the original size.
    The last convolution layer outputs the probability of the
    seven signal labels at each pixel.
    The probability of each label is rounded to a binary value.
    Since the activation in the last layer is a sigmoid function,
    which does not normalize the output probability,
    the model retains the capability of predicting multiple labels
    for a same pixel.
    The drop rate of the DropBlock in the original SA-UNet is calculated
    as follows:
    \begin{linenomath*}
    \begin{eqnarray}
        \gamma = \frac{1 - P}{K^2}\frac{H}{H - K + 1}\frac{W}{W - K + 1},
    \end{eqnarray}
    \end{linenomath*}
    where $\gamma$ is the drop rate,
    $K$ is the kernel size, $H$ is the height of the image,
    $W$ is the width of the image, and $P$ is the probability to keep the 
    kernel.
    To reduce the computational complexity, we slightly modified the 
    above formula of drop rate to the following:
    \begin{linenomath*}
    \begin{eqnarray}
        \gamma = \frac{1 - P}{K^2}.
    \end{eqnarray}
    \end{linenomath*}

    \subsection{Model Training, Evaluation and Optimization}
    In each training epoch, a mini-batch is fed into the model,
    and the loss is obtained by calculating the binary crossentropy between the
    ground truth and the prediction.
    The model weights are then updated by backpropagating the gradients 
    obtained by the AMSGrad optimizer \cite{amsgrad}.
    An epoch is completed after all mini-batches are used up.
    The learning rate is set to $0.001$ initially, and is halved until it 
    reaches a minimum rate of $10^{-6}$ if no improvement of the loss of the 
    validation data is found from the previous $5$ epochs.
    A final model is selected from all the epochs by manually checking the
    performance of the validation set.
    We found no significant improvement after $60$ epochs.

    We measure the model performance by the intersection over union (IoU).
    We modify the definition of IoU in order to apply this to $7$ labels.
    The IoU of each label $k$ in an individual ionogram $n$,
    $\mathrm{IoU}_{n,k}$ is computed as the ratio of the total number of
    intersecting pixels of the ground truth and the prediction
    ($\mathrm{I}_{n,k}\equiv\mathrm{truth}\cap\mathrm{prediction}$)
    over the total number of union pixels of ground truth and prediction
    ($\mathrm{U}_{n,k}\equiv\mathrm{truth}\cup\mathrm{prediction}$).
    The mean IoU of each label $k$, $\mathrm{IoU}_{k}$, is computed as the
    sum of the intersecting pixels over the sum of the union pixels of all
    ionograms.
    The equations for IoU$_{n,k}$ and mean IoU$_{k}$ are as follows:
    \begin{linenomath*}
    \begin{eqnarray}
        \mathrm{IoU}_{n, k} &=& \frac{\mathrm{I}_{n,k}}{\mathrm{U}_{n,k}},\\
        \textrm{mean IoU}_k &=& \frac{\sum_{n=1}^{N}\mathrm{U}_{n,k}\cdot
        \mathrm{IoU}_{n,k}}{\sum_{n=1}^{N}\mathrm{U}_{n,k}}, \nonumber \\
        &=& \frac{\sum_{n=1}^{N}\mathrm{I}_{n,k}}{\sum_{n=1}^{N}\mathrm{U}_{n,k}}.
    \end{eqnarray}
    \end{linenomath*}
    The mean IoU is calculated in such a way that we do not
    encounter zero-division for labels with both $I_{n,k}=0$ and $U_{n,k}=0$.
    An example of application of this technique is shown in
    Figure~\ref{fig:echo_ffp_fp} (see Section~\ref{sec:recover}).

    The model is optimized by tuning the hyperparameters.
    In this study, the hyperparameter includes the size of the convolution 
    kernel, the keep probability of DropBlock, the size of the mini-batch.
    The default hyperparameters are kernel size of $3\times3$,
    keep probability of $0.85$, and batch-size of $4$.
    To prevent a large searching grid, the hyperparameter is tuned individually
    while the others are fixed to their default values.
    The size of the convolution kernel varies between $3\times3$, $5\times5$,
    and $7\times7$;
    the keep probability varies between $0.5$, $0.7$, and $0.85$;
    the size of the mini-batch varies between $1$, $2$, $4$, and $8$.
    This results in $10$ different combinations of hyperparameters.
    By manually comparing the IoUs of the validation data from the $10$ 
    combinations of hyperparameters,
    we determine the optimal hyperparameters as kernel sizes of
    $7\times7$, and the batch size of $4$.
    We found the optimal performance occurs at the $44$th epoch.

    \section{Result}
    \label{sec:result}
    \subsection{Recovery of Ionogram Signals}
    \label{sec:recover}

\begin{figure}
        \centering
        \includegraphics[width=.9\textwidth]{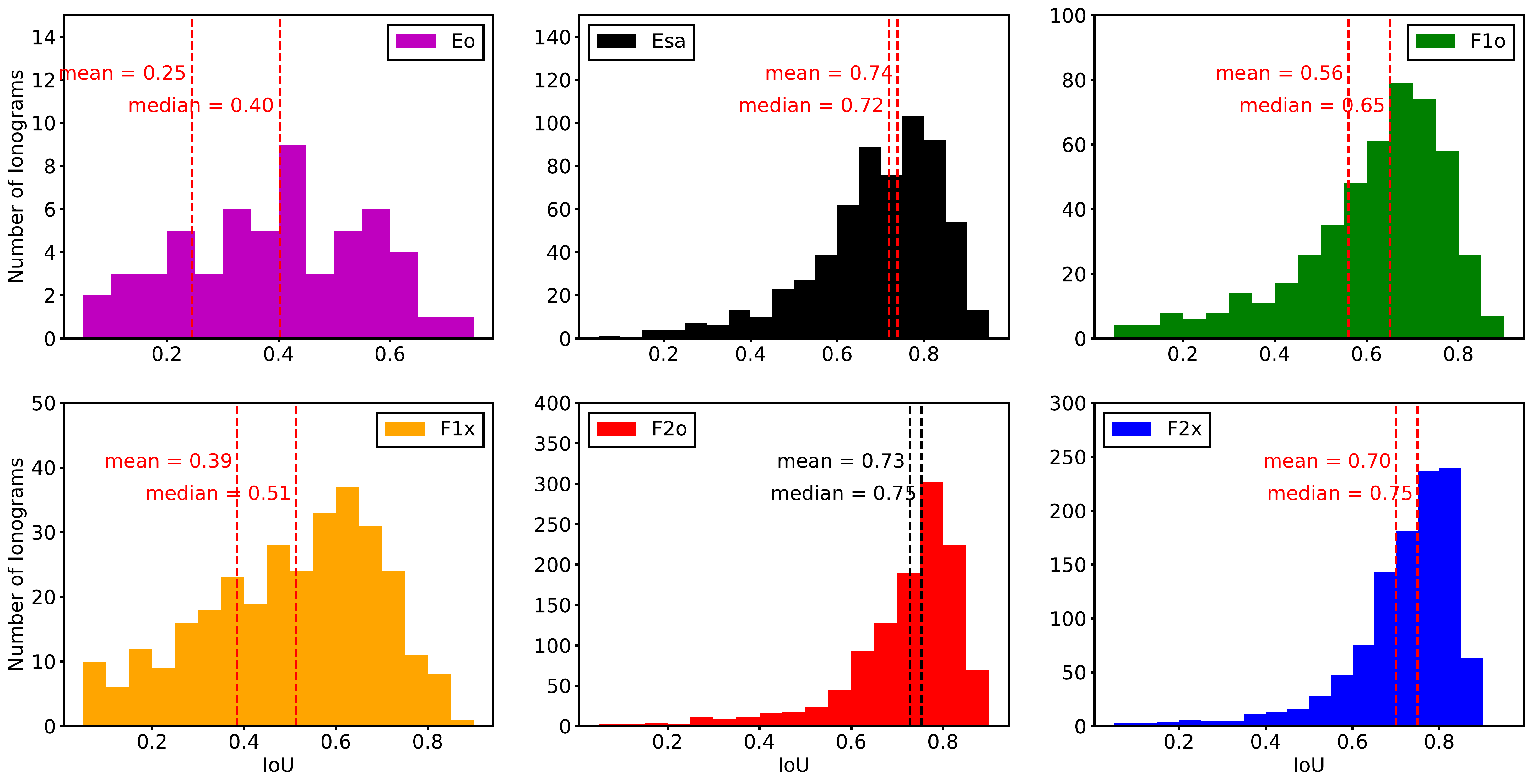}
        \caption{Distribution of non-zero IoU of each label, and the 
        corresponding mean IoU and median IoU.
        The vertical dashed lines in each panel mark the locations of
        the mean IoU and median IoU.
        }
        \label{fig:iou_dist}
    \end{figure}

    The distribution of IoUs of each label in the test set is plotted
    in Figure~\ref{fig:iou_dist} with the mean IoU printed in the corresponding
    panels.
    The figure shows that the most accurately recovered signals by
    SA-UNet are Esa, F2o, and F2x, with mean IoUs reaching approximately $0.7$.
    F1o and F1x are less well recovered, with mean IoU equal to
    $0.56$ and $0.39$, respectively.
    The near-zero mean IoU of Eo and Ex
    indicates that the model cannot identify these two signals.
    Comparing the mean IoUs of different labels and their coverage ratio in
    the training set (Table~\ref{tab:datacov}) indicates that
    the two are highly related:
    The labels with coverage ratio over $50\%$ are the best recovered
    labels, and the two labels, Eo and Ex, with the lowest coverage ratios
    are the worst recovered.

    A notable amount of zero IoUs are observed.
    We divide these zero IoUs into three different types
    (see Table~\ref{tab:zero_ious}).

    \begin{table}
        \centering
        \begin{tabular}{*{8}{c}}
            \hline
             & Eo & Ex & Esa & F1o & F1x & F2o & F2x\\
            \hline
            Test set & 83 & 10 & 638 & 490 & 339 & 1160 & 1083\\
            Zero IoUs & 52 & 10 & 36 & 114 & 209 & 22 & 88\\
            FN & 28 & 10 & 14 & 3 & 11 & 3 & 1\\
            FFP & 17 & 0 & 9 & 16 & 95 & 9 & 52\\
            FP & 7 & 0 & 13 & 95 & 103 & 10 & 35\\
            \hline
        \end{tabular}
        \caption{Rows from top to bottom are the number of instances in 
        the test set, the number of instances with zero IoUs, 
        the number of correctly labeled signals not identified by the model
        (FN); unlabled signals correctly identified by the model (FFP);
        and incorrectly predicted signals (FP).}
        \label{tab:zero_ious}
    \end{table}

\begin{figure}
        \centering
        \includegraphics[width=\textwidth,clip,trim=0 6.5cm 0 4.5cm]{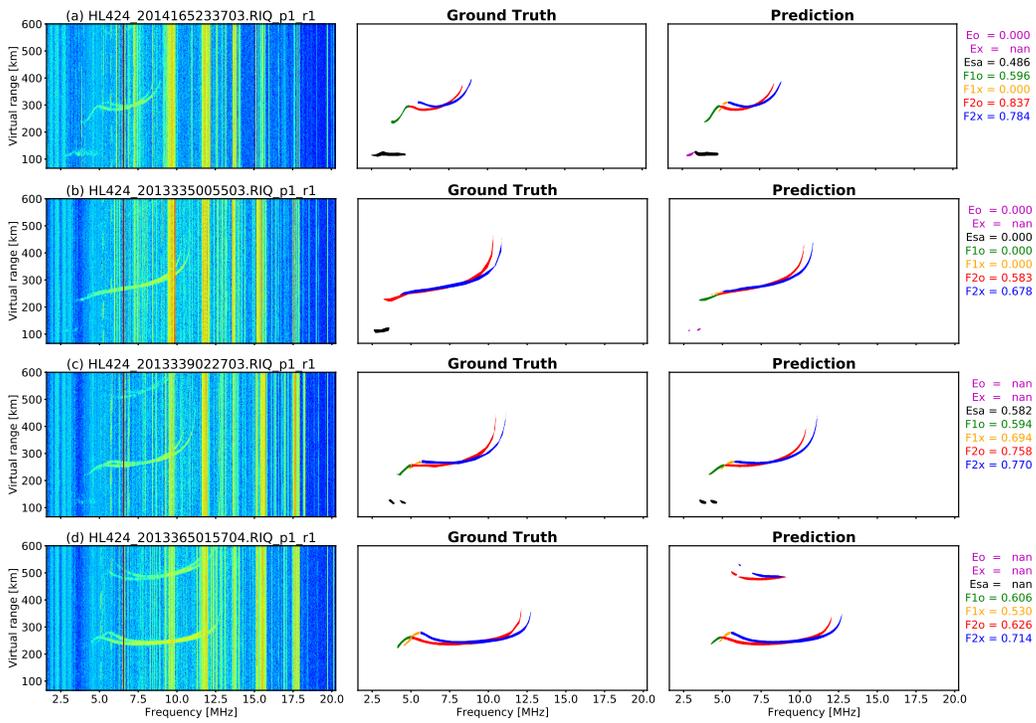}
        \caption{
        The panels from left to right are
        the ionograms, their corresponding ground truths, and 
        model predictions.
        The IoU of each label is printed next to the panels of model prediction.
        The measurement time of each ionogram: 
        (a) 2014/06/14 23:37:03, (b) 2013/12/01 00:55:03, 
        (c) 2013/12/05 02:27:03, and (d) 2013/12/31 01:57:04.
        }
        \label{fig:echo_ffp_fp}
    \end{figure}

    \begin{enumerate}
    \item
    Model fails to identify correctly labeled signals (false negative, FN):
    This happens when the model either fails to identify the signal or incorrectly identities it as other labels.
    Such situation commonly occurs for the signals with low coverage ratio,
    which can result in model being undertrained.
    In fact, all Ex instances in $10$ ionograms are identified as Eo label.

    \item
    Model correctly identifies the signals that are not labeled
    (false false-positive, FFP):
    This happens when the model correctly identifies the signals that 
    were not labeled or incorrectly labeled.
    The signals that are overlapping with other signals
    or are contaminated by strong noise
    are difficult for human to accurately label them.
    The correct identification of such signals by our model indicates its
    superior capability over human eye.
    One such example can be seen in Figure~\ref{fig:echo_ffp_fp}a:
    there is a tiny Eo signal in the ionogram
    that was incorrectly labeled as Esa due to strong noise.
    The model correctly separates the signal into increasing (Eo) and 
    the decreasing (Esa) part, 
    resulting in zero IoU for Eo and lower IoU for Esa.
    \item
    Model incorrectly identifies the signals that do not exist
    (false positive; FP):
    Such situation occurs when the model is overtrained to a specific label,
    thus trying to identify too many unrelated pixels as the label.
    For example, in panel (b) of Figure~\ref{fig:echo_ffp_fp}, 
    there are only F2o and F2x signals labeled in the ground truth
    data, but the model divides the beginning parts of F2o and F2x as
    F1o and F1x signals.
    \end{enumerate}

    In addition to different types of zero IoUs, mean IoU can also be
    decreased by echo signals being identified as the primary ones.
    Echo signals are the signals bouncing between the ground and the
    ionosphere.
    In the ionograms, they have similar shapes as the main signals,
    but appear at higher altitudes with weaker amplitudes.
    As shown in panel (c) of Figure~\ref{fig:echo_ffp_fp}, the model correctly 
    identifies the signals from the ionosonde measurement.
    However, in panel (d), the echo signals at higher altitude are also
    identified by the model as F2o and F2x labels,
    lowering their IoUs as a result.

\begin{figure}
        \includegraphics[width=.9\textwidth,clip,trim=0 1cm 3.5cm 1.5cm]
        {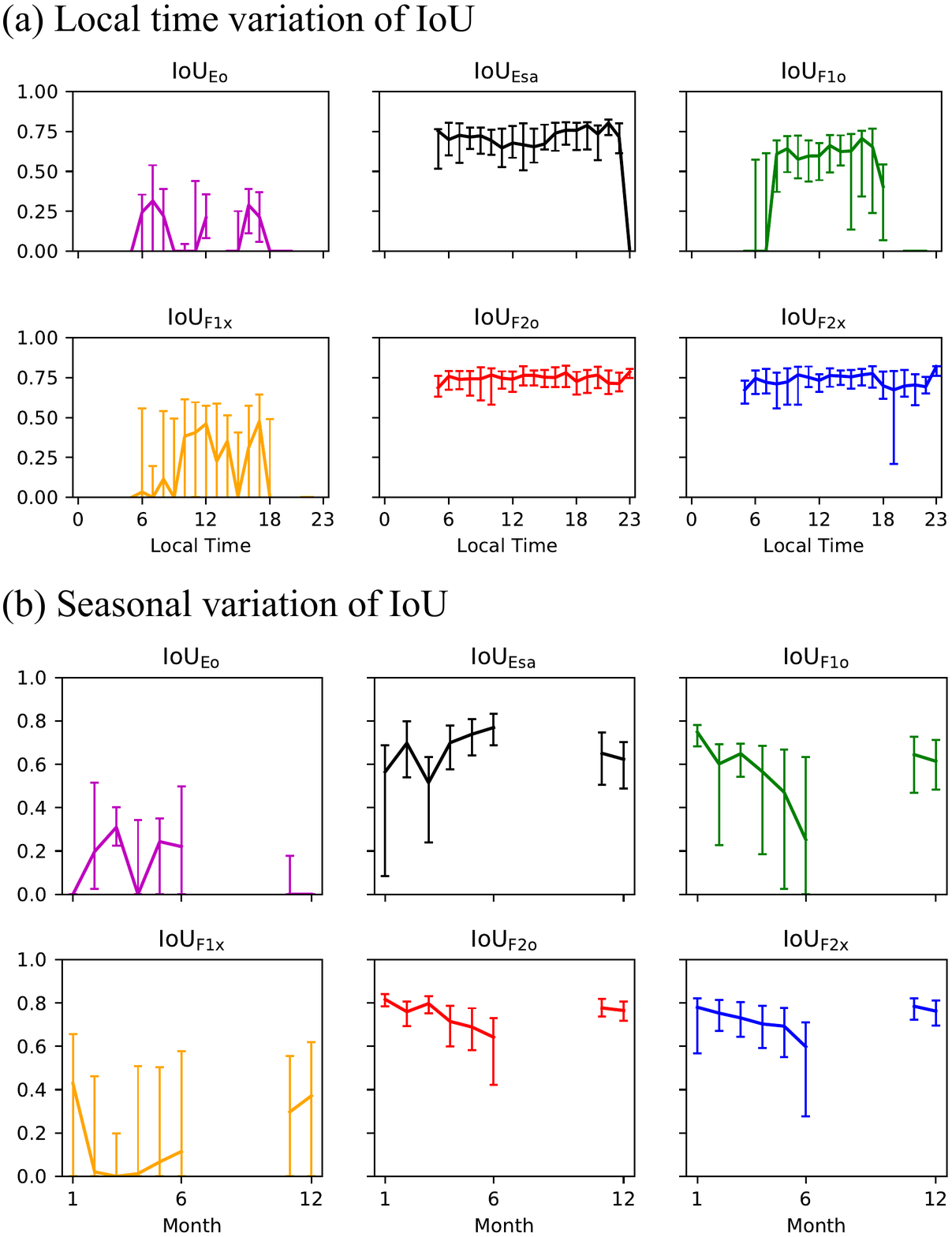}
        \caption{The panels in (a) and in (b) are the local time and seasonal
        variations of IoUs, respectively, of the 7 signal labels, as
        indicated above the corresponding panels. All panels follow the scale
        of the left most y-axis.}
        \label{fig:iou_lt}
    \end{figure}

    To investigate the local time and the seasonal dependency of the model
    performance,
    we consider the median value of IoU, and use the first and the third
    quartiles as the error bar.
    The panels in Figure~\ref{fig:iou_lt}a show the local time variation 
    of IoUs of Eo, Esa, F1o, F1x, F2o, and F2x of the test.
    Apart from 23:00, the median value of IoU of
    Esa, F2o, and F2x labels in general are independent of the local time.
    The panels in Figure~\ref{fig:iou_lt}b show the seasonal variation of the 
    IoUs.
    The IoUs of F1 (F2o) and F2 (F2o and F2x) layers have a tendency
    to decrease in the summer, while the IoU of Esa label increases.
    This could be caused by non-uniform statistics of the layers.
    Namely, sporadic Es has higher occurrence and intensity during
    summer months.
    Strong Es layer very often hide the F2 layer such that the statistics of 
    those labels decrease in summer \cite{snr}.

    \subsection{Examination of Ionogram Parameters}
    \label{sec:param}
    From the ionograms recovered by the deep learning model,
    we extract the virtual heights, the critical frequencies and
    the intersection frequencies by an automated procedure.
    In this study, we only present the parameters for F2 signals,
    because our model performs the best on F2 signals.
    The same analysis can also be applied to other signals.
    For a signal X $\in\left\{\mathrm{F2o, F2x}\right\}$ in the
    $n$-th recovered ionogram, $X_n(f_i, h'_j)$ at frequency $f_i$ and
    virtual height $h'_j$ is defined as:
    \begin{linenomath*}
    \begin{equation}
        X_{n}(f_i, h'_j) = \left\{
        \begin{array}{l}
            1,\;\mathrm{signal\ of\ X}; \\
            0,\;\mathrm{otherwise.}
        \end{array}
        \right.
    \end{equation}
    \end{linenomath*}

\begin{figure}
        \centering
        \includegraphics[width=\textwidth,clip,trim=0 2.5cm 0 2.5cm]
        {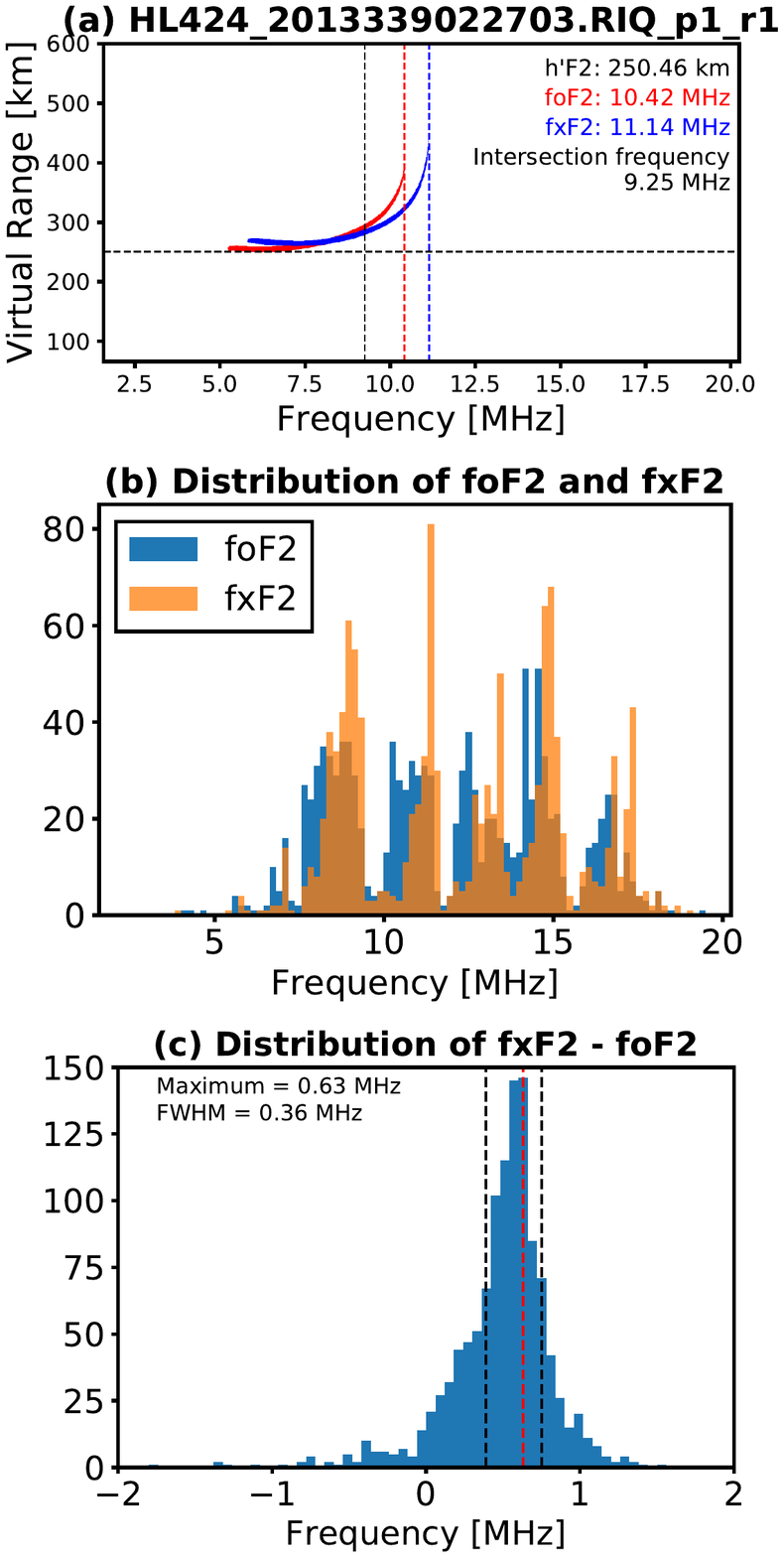}
        \caption{(a) The recovered ionogram at 2013/12/05 02:27:03 showing
        only F2o and F2x signals.
        The corresponding critical frequencies of F2o and F2x, and the
        intersection frequency between the two signals are $10.42$ MHz,
        $11.14$ MHz, and $9.25$ MHz, respectively, marked by the red, blue,
        and the black dashed line.
        (b) The distribution of foF2 (blue bars) and fxF2 (orange bars).
        (c) The distribution of fxF2 $-$ foF2. Red dashed line indicates
        the location of the maximum, and black dashed lines indicate the
        location of the half-maximum.}
        \label{fig:intersect}
    \end{figure}

    The virtual height of the F2 signal is defined as the minimum of $h'_j$
    where F2o$(f_i, h'_j)$ is non-zero:
    \begin{linenomath*}
    \begin{eqnarray}
        \mathrm{h'F2}_n &=&
        \min h'_j\;\mathrm{of}\;\mathrm{F2o}_n(f_i, h'_j), \\
    \end{eqnarray}
    \end{linenomath*}
    and the critical frequencies
    foF2 and fxF2 are defined as the maximum of $f_i$ 
    at which F2o$(f_i, h'_j)$ and F2x$(f_i, h'_j)$ are non-zero:
    \begin{linenomath*}
    \begin{eqnarray}
        \mathrm{foF2}_n &=& \max f_i\;\mathrm{of}\;\mathrm{F2o}_n(f_i, h'_j), \\
        \mathrm{fxF2}_n &=& \max f_i\;\mathrm{of}\;\mathrm{F2x}_n(f_i, h'_j),
    \end{eqnarray}
    \end{linenomath*}
    and the intersection frequency between the
    F2o and the F2x signals, is defined as the first intersection point
    from the high frequency.
    \begin{linenomath*}
    \begin{equation}
        \mathrm{Intersection\ frequency(F2o, F2x)} = 
        \max{f_i \; \mathrm{ of } \; (\mathrm{F2o}_n \cap \mathrm{F2x}_n)}
    \end{equation}
    \end{linenomath*}
    For F2o and F2x signals with large overlapping area,
    the extracted intersection frequency is not reliable and cannot be used.
    We consider the overlapping area greater than $0.3$ of their combined area
    as highly overlapped.
    Out of $1226$ ionograms recovered from the test set,
    $702$ are below the threshold.
    One example is shown in Figure~\ref{fig:intersect}a.
    The corresponding original ionogram is measured at 2013/12/05 02:27:03,
    the same timestamp as the ionogram in Figure~\ref{fig:echo_ffp_fp}c.
    For this timestamp, the virtual height of F2 is $250.46$ km,
    the critical frequencies foF2 and fxF2 are $10.42$ MHz,
    and $11.14$ MHz, respectively, and the intersection frequency between
    the F2o signal and the F2x signal is $9.25$ MHz.

    The distributions of foF2 and fxF2 are plotted in
    Figure~\ref{fig:intersect}b.
    It shows that the extracted foF2 varies from $4.07$ MHz to $19.36$ MHz.
    The distribution of the difference between fxF2 and foF2 (fxF2$-$foF2)
    is shown in Figure~\ref{fig:intersect}c.
    It shows a distribution close to a Gaussian profile, with a tail on the
    left-hand side.
    The peak of the distribution is $0.63$ MHz, and the full-width half-maximum
    (FWHM) is $0.36$ MHz, making the uncertainty of the difference (FWHM$/2$)
    to be $0.18$ MHz.

    \section{Discussion}
    \label{sec:discuss}
    Studies have shown that the deep learning models are able to scale the
    ionograms automatically.
    The reported performance of the deep learning models are summarized
    and compared with our result in Table~\ref{tab:ious_model}.
    \citeA{aeperu} applied the Autoencoder model (denoted by the superscript a)
    to extract F region signals
    from Peru's ionograms, and obtained IoU$\approx 0.6$ for the combined
    F layer signals (F1 $+$ F2), after fine-tuning the model parameters.
    The Fully Convolutional DenseNet model (denoted by the superscript b)
    used in \citeA{snr} to recover
    the ionospheric signals obtained an IoU of nearly $0.6$ for Esa and
    F2o layers.
    \citeA{unetseg} used multiple U-Net models (denoted by the superscript c)
    to scale E, F1, F2 layers,
    and obtained the dice-coefficient loss (DCL) of $0.16$, $0.17$, and $0.11$
    for E, F1 and F2 layers, respectively, in the ionograms of 64x48 pixels,
    and $0.22$, $0.23$, and $0.19$ in the ionograms of 192x144 pixels.
    Note that a smaller DCL indicates a better performance.
    Our SA-UNet model, applied to highly noisy ionograms of 800x1600 pixels,
    achieves an IoU $\ge 0.7$ and DCL score $\le 0.18$ for Esa and F2 layers.
    To provide additional information for interested readers, 
    we also list the recall rates and precision rates achieved 
    by our model, which are calculated
    by counting the pixel number of true positives, false positives, and
    false negatives.
    It should be noted that the ionograms 
    used by \citeA{aeperu} and \citeA{unetseg}
    have been obtained either at middle latitudes or/and in the remote regions
    with low man-made signals.
    In short, our study shows that the SA-UNet can improve the performance to
    IoU $\approx 0.7$ and reduce DCL to below 0.18.

    \begin{table}
        \centering
        \begin{tabular}{*{10}{c}}
            \hline
             & Size (HxW) & Metric & Eo & Ex & Esa & F1o & F1x & F2o & F2x\\
            \hline
            Autoencoder$^a$ & 256x208 & IoU & N/A & N/A & N/A
                            & \multicolumn{4}{c}{0.60} \\
            \hline
            FC-DenseNet$^b$ & 800x1600 & IoU & 0.00 & 0.00 & 0.56 & 0.47 & 0.33 & 0.59 & 0.48\\
            \hline
            U-Net$^c$ & 64x48 & DCL & \multicolumn{3}{c}{0.16}
                      & \multicolumn{2}{c}{0.17} & \multicolumn{2}{c}{0.11}\\
                      & 192x144 & DCL & \multicolumn{3}{c}{0.22}
                      & \multicolumn{2}{c}{0.23} & \multicolumn{2}{c}{0.19}\\
            \hline
            SA-UNet & 800x1600 & IoU & 0.25 & 0.00 & 0.74 & 0.56 & 0.39
                    & 0.73 & 0.70\\
                    & & Recall & 0.28 & 0.00 & 0.85 & 0.73 & 0.51
                               & 0.83 & 0.83 \\
                    & & Precision & 0.68 & N/A & 0.85 & 0.71 & 0.61 
                                  & 0.86 & 0.82 \\
                    & & DCL & 0.61 & 1.00 & 0.15 & 0.28 & 0.44 
                            & 0.16 & 0.18 \\
            \hline
        \end{tabular}
        \caption{The reported performance of different models, and the 
        corresponding size of the input ionograms in units of pixels.
        DCL denotes the dice-coefficient loss.
        All metrics are calculated pixel-wise.}
        \label{tab:ious_model}
    \end{table}

    The critical frequency foF2 is directly related to the maximum number
    density $N$ of electrons in F2 layer, and the difference between
    fxF2 and foF2 can provide an estimation of the local geomagnetic field
    $B$ \cite{handbook}:
    \begin{linenomath*}
    \begin{eqnarray}
        N &=& 1.24\times10^{10}\times(\mathrm{foF2} / \mathrm{MHz})^2
        \;[\mathrm{m}^{-3}] \\
        B &\approx& 0.71\times(\mathrm{fxF2} - \mathrm{foF2})\;[\mathrm{G}].
    \end{eqnarray}
    \end{linenomath*}

    Our results show that the derived electron number density ranges from
    $1.66\times10^{11}$ m$^{-3}$ to $3.75\times10^{12}$ m$^{-3}$,
    and the magnetic field is $B \approx 0.45\pm0.13$ G.
    Since the local variation of the geomagnetic field is usually small,
    the large uncertainty of the magnetic field is likely caused by strong
    noises in the ionograms.
    We also find 76 recovered ionograms with foF2 $>$ fxF2.
    They are caused by both false positive and false negative predictions.

    The existence of echoes does not affect the extraction of
    critical frequencies.
    However, cares should be taken for extracting the critical
    virtual heights.
    In addition to the local geomagnetic field,
    the difference between foF2 and fxF2 can also be used to estimate
    foF2 (i.e. $\mathrm{foF2} \approx \mathrm{fxF2} - 0.63\,\mathrm{MHz}$)
    in the case when fxF2 is obtained but F2o cannot be accurately recovered.

    The intersection frequency can be used for the verification of the ordinary
    and extra-ordinary signals from the F2 layer.
    Namely, the ascending branch of the ordinary signal should be situated
    at lower frequencies than the extra-ordinary one.
    This technique can provide a robust correction of inaccurate model
    predictions of the signals from the F2 layer and, thus,
    more accurate determination of the critical frequencies foF2 and fxF2.

    The low percentages of 
    Eo and Ex labels in the training, validation and the test set 
    could bias the model performance.
    In order to generalize the data representation,
    an k-fold or a leave-one-out cross validation may be applied \cite{cv1,cv2}.
    Moreover, there are a few techniques which may reduce the 
    class imbalance problem, such as undersampling/oversampling to the 
    training set \cite{ousampling},
    loss weighted according to the amount of the labels \cite{wtloss1,wtloss2},
    transfer learning from a larger dataset, or ensemble learning of multiple 
    models, to name a few.

    It is important to note that the present dataset spans less than
    one year, and includes only ionospheric statistics at low latitudes
    under very dynamic region of the bulge of equatorial ionization anomaly.
    The ionosphere dynamics depends on latitude and varies with solar and 
    geomagnetic activity.
    Hence, to enable the application of our model to other datasets,
    the model should be trained on the combination of different data set.
    Alternatively,
    the transfer learning technique can be applied to reduce
    the training time and improve the model performance on a different data set.

    \section{Conclusions}
    \label{sec:summary}
    In this study, we show that SA-UNet is capable of recovering
    F2o and F2x signals, and the Esa label (which is a combination of
    Eso, Esx, and Es signals) from the highly contaminated Hualien VIPIR
    measurements.
    While the performance of identifying Eo and Ex 
    is poor due to the class imbalance, their low percentage 
    in the statistics suggests that they have little effect 
    on the ionosphere in the Taiwan region.
    By comparing the input ionogram, ground truth labeling,
    and the model prediction, our results show that the model is capable of 
    recovering signals that are not labeled, due to contamination by strong 
    noise or overlapping with another signal.
    The recovered ionograms can be further used in extracting the
    virtual height, the critical frequency and the intersection frequency of 
    different labels, which are important in the determination of the electron 
    density profile and the magnetic field of the ionosphere overhead.

    \section{Open Research}
    The ionogram data and the model used in this study can be openly accessed on
    Kaggle.
    The link for the data is provided as follows:
    \url{https://www.kaggle.com/changyuchi/ncu-ai-group-data-set-fcdensenet24}.
    The link for the model is provided as follows:
    \url{https://www.kaggle.com/guanhanhuang/ncu-ai} \cite{ncuai}.
    The model is constructed using Tensorflow 2.3 \cite{tensorflow} and
    its Keras packages,
    and is trained on Kaggle under the GPU computation environment.

    \acknowledgments
    This work is funded by the Ministry of Science and Technology of Taiwan
    under the grant number 109-2923-M-008-001-MY2 and 109-2111-M-008-002.
    The figures are made by using the Matplotlib package \cite{matplotlib}.

    \bibliography{refs1}

\end{document}